\documentclass[a4paper,11pt]{article}

\usepackage[english]{babel} 
\usepackage[utf8x]{inputenc} 
\usepackage[T1]{fontenc}

\usepackage[a4paper,top=2.5cm,bottom=2.5cm,left=2.5cm,right=2.5cm,marginparwidth=1.75cm]{geometry}

\usepackage{amsmath,subcaption,amsfonts,amssymb}
\usepackage{siunitx}
\usepackage{graphicx,blkarray} 
\usepackage[colorinlistoftodos]{todonotes} 
\usepackage[colorlinks=true, allcolors=blue]{hyperref} 
\usepackage[numbers,sort&compress]{natbib}
\usepackage[onehalfspacing]{setspace}
\mathchardef\mhyphen="2D
\date{}
\title{Evolution of cooperation on an epithelium} 
\author{Jessie Renton\thanks{Corresponding author: jessica.renton.16@ucl.ac.uk} \qquad Karen M.\ Page \\ 
\emph{Department of Mathematics, University College London,}\\\emph{Gower Street, London WC1E 6BT, UK}}
\begin{document} 
\maketitle
\begin{abstract}
	Cooperation is prevalent in nature, not only in the context of social interactions within the animal kingdom, but also on the cellular level. In cancer for example, tumour cells can cooperate by producing growth factors. The evolution of cooperation has traditionally been studied for well-mixed populations under the framework of evolutionary game theory, and more recently for structured populations using evolutionary graph theory. The population structures arising due to cellular arrangement in tissues however are dynamic and thus cannot be accurately represented by either of these frameworks. In this work we compare the conditions for cooperative success in an epithelium modelled using evolutionary graph theory, to those in a mechanical model of an epithelium- the Voronoi tessellation model. Crucially, in this latter model cells are able to move, and birth and death are not spatially coupled. We calculate fixation probabilities in the Voronoi tessellation model through simulation and an approximate analytic technique and show that this leads to stronger promotion of cooperation in comparison with the evolutionary graph theory model.	
\end{abstract}

\begin{small} \textbf{\emph{Keywords--- }} cooperation, epithelium, population structure, evolutionary graph theory, Voronoi tessellation
\end{small}
\section{Introduction} 
\label{sec:introduction}

Tumour development is an evolutionary process whereby cells undergo a series of genetic changes leading to acquired capabilities that confer some growth advantage. In Hanahan and Weinberg's seminal paper \cite{Hanahan2000}, six such capabilities or `hallmarks of cancer' were identified to be necessary for normal cells to become malignant: self-sufficiency in growth signals, insensitivity to antigrowth signals, evading apoptosis, limitless replicative potential, sustained angiogenesis, and tissue invasion and metastasis. Many of these rely on the production of diffusible growth factors \cite{Witsch2010}, the effects of which are felt not only by the producer cell but by other cells in the neighbourhood. As such production of these growth factors can be considered an example of cellular cooperation \cite{Jouanneau1994,Axelrod2006}. 

The reprogramming of energy metabolism is also considered to be a hallmark of cancer \cite{Hanahan2011} and refers to the fact that cancer cells tend to metabolise through glycolysis rather than aerobic respiration, even when oxygen is abundant. This is known as the Warburg effect \cite{Warburg1956}. It has been postulated that glycolytic cells can be considered as cooperators, in that they produce lactic acid as a by-product which is toxic to healthy cells, and can thus be considered a shared benefit to the cancer cells. 

Models of the evolution of cooperation for diffusible growth factors \cite{Tomlinson1997,Bach2001,Archetti2013b} and the Warburg effect \cite{Basanta2008,Basanta2011,Kareva2011} have been developed using the framework of evolutionary game theory for well-mixed populations. These models have also been extended to consider spatial effects by placing cells on a lattice \cite{Bach,Archetti2013a,Archetti2013} or a fixed graph \cite{Archetti2015,Archetti2016}. Further examples of the application of game theory to cancer evolution include microenvironment dependency \cite{Anderson2009,Basanta2012}, environmental poisoning \cite{Tomlinson1997a} and invasion \cite{Basanta2008a}. See \cite{Hummert2014} for a recent review of evolutionary game theory applied to somatic evolution.

Cell populations are not well-mixed but organised into tissues or tumours, thus the recent move to incorporate spatial structure is important. Introducing population structure can have a significant effect on evolutionary dynamics \cite{Nowak2010}, in particular in promoting cooperation \cite{Nowak2006}. The established framework for modelling games on structured populations, used in the models mentioned above, is evolutionary graph theory (EGT) \cite{Lieberman2005,Ohtsuki2006,Ohtsuki2006b,Szabo2007,Taylor2007,Ohtsuki2008,Allen2016} in which individual cells are placed on the vertices of a graph and neighbours are joined together by edges. Individuals interact and play games with their neighbours, thus deriving their fitnesses. The population evolves via some update rule which dictates how birth and death occur while maintaining the fixed graph structure. When a cell divides it is necessary for a neighbouring cell to die in order that one of the offspring can occupy the empty vertex. Two commonly used update rules are the birth-death and death-birth rules which essentially differ in the order in which birth and death events occur.

There are several shortcomings of evolutionary graph theory in application to somatic evolution. Tissue and tumour structures are not fixed but dynamic, due to processes of cell division, extrusion and motility. Furthermore the necessity of births and deaths occurring next to each other is not only unrealistic, but the choice of update rule is one of the main determinants of evolutionary outcomes \cite{Zukewich2013}. Recent work has introduced a new `shift update' with the aim of addressing the unsuitability of the traditional update rules for cellular structures. The model works extremely well in one-dimension \cite{Allen2012}, predicting enhanced cooperative success compared to other update rules. However the extension into two-dimensions \cite{Pavlogiannis2015} is not straightforward as the shifting of cells disrupts cluster formation of cooperators. This can be resolved by introducing a repulsive force between cells of different types and choosing energy-minimising shift paths. If the force is strong enough the shift dynamics is again an effective promoter of cooperation. However it relies on this somewhat artificial preferential sorting.

Dynamic graph models of evolutionary games also exist, however they focus on switching connections between vertices, either at random or to increase fitness \cite{Pacheco2006,Santos2006,Wu2010,Pinheiro2016}. These types of models are relevant in social networks, for example, where agents can choose who they interact with and can break social ties with individuals who do not cooperate \cite{Moreira2013}. They are not good models, however, for populations of cells which are spatially constrained in two- or three-dimensional structures. Furthermore they still require birth and death to be coupled.

In order to elucidate what impact, if any, the dynamic nature of cell populations and spatial decoupling of birth and death has on the evolution of cooperation, we will consider evolutionary games on a mechanical model of an epithelium- the Voronoi tessellation (VT) model \cite{Meineke2001,VanLeeuwen2009}. Epithelia are the tissues which form the surfaces in the body, such as skin, and the linings of organs. We choose this particular tissue structure as it can be modelled in two dimensions as a sheet of polygonal cells \cite{Osborne2017}. Furthermore epithelial cells are highly proliferative compared to other cell types and the source of 85\% of cancers making them of particular interest in models of cancer evolution. 


Rather than focussing on a particular cancer model, we consider the simple, and well studied, example of an additive prisoner's dilemma game, whereby cooperators pay a cost $c$ in order to produce some benefit $b$ for their neighbours. We consider whether the results for the VT model are significantly different from those obtained from EGT. In particular we calculate the fixation probabilities for single mutant cooperators arising in a population of defectors in both models.

We begin, in Section~\ref{sec:evolutionary_graph_theory}, by introducing EGT and looking at how it can be applied to the evolution of cooperation on epithelia, considering results for an additive prisoner's dilemma game with both birth-death and death-birth update rules. We then, in Section~\ref{sec:VTmodel}, introduce the VT model of an epithelium, again considering the evolution of cooperation under a prisoner's dilemma, but this time with spatially decoupled birth and death. We calculate approximate fixation probabilities as well as looking at simulation results. Finally in Section~\ref{sec:comparing_the_models} we compare these results with the EGT model, finding that cooperation is significantly more successful in the VT model. By running further simulations, implementing an explicit death-birth update in the VT model and a migration analogue into the EGT model, we identify the decoupling of birth and death to be the primary mechanism for the discrepancy.


\section{Evolutionary graph theory} 
\label{sec:evolutionary_graph_theory}
\subsection{The model} 
\label{sub:model_EGT}

Evolutionary graph theory provides a framework for modelling the evolution of traits on fixed population structures represented by a static graph $G$. Individuals, labelled $i=1,2,...,N$ for a population size $N$, are represented by the vertices, while the edges correspond to neighbour connections. We therefore define the adjacency matrix 
\begin{equation}
	A_{ij} = 
	\begin{cases}
		1, &\text{ if $i$ and $j$ are neighbours}\\
		0, &\text{ otherwise.} 
	\end{cases}
\end{equation}
In the additive prisoner's dilemma, the trait or type of an individual $i$ is given by $s_i\in\{0,1\}$, with $s_i=0$ denoting a defector (D) and $s_i=1$ a cooperator (C). The state of the population is then given by the $N$-dimensional vector $\mathbf{s}$.

For a population in state $\mathbf{s}$, individual $i$ obtains a payoff $f_i(\mathbf{s})$ from its neighbours which is calculated according to a payoff matrix, given by
\begin{equation}
	\begin{blockarray}
		{ccc} & C & D \\
		\begin{block}
			{c(cc)} C & b-c & -c \\
			D & b & 0 \\
		\end{block}
	\end{blockarray} \; ,
\end{equation}
where $b>c$ and $c>0$. The payoffs are thus
\begin{equation}
	f_i\mathbf{(s)} = -cs_i + b\sum_{j\in G}\frac{A_{ij}s_j}{k_i} \label{eq:payoffs} \; ,
\end{equation}
where $k_i=\sum_{j\in G}A_{ij}$ is the degree of vertex $i$ (i.e.\ the neighbour number). Fitness is then defined to be
\begin{equation}
	F_i\mathbf{(s)}=1+\delta f_i(\mathbf{s}) \; ,
\label{eq:fitness} \end{equation}
where $\delta>0$ is the selection strength parameter and the constant $1$ takes into account other contributions to fitness. We can let $c=1$ without loss of generality, thus the game is defined by a single parameter.

Evolution proceeds via a spatial extension of the Moran process \cite{Moran1958,Lieberman2005} whereby, at each time step, an individual dies and another reproduces. The offspring occupies the vacant vertex thus keeping the graph structure constant. There are several potential mechanisms for this, known as update rules. Here we consider two common rules: 
\begin{itemize}
	\item \emph{birth-death:} an individual is chosen to reproduce with probability proportional to fitness; its offspring takes the site of a neighbour selected uniformly at random to die; 
	\item \emph{death-birth:} an individual is chosen to die uniformly at random; it is replaced by the offspring of a neighbour chosen with probability proportional to fitness. 
\end{itemize}
For a well-mixed population, represented by a complete graph, these two updates rules are equivalent, however for an arbitrary population structure the choice of update rule leads to strikingly different dynamics. In the following we will consider the dynamics in both cases for graph structures representing an epithelium.

\subsection{Fixation probabilities} 
\label{sec:games_on_epithelium}
In order to consider game dynamics on an epithelium within the EGT context we consider two different graph structures. Epithelial cells have six neighbours on average, therefore a hexagonal lattice (HL) is a simple approximation. A Voronoi tessellation however gives a more realistic representation of an epithelium \cite{Honda1978,Zhu2001,Sanchez-Gutierrez2015}. There is some variance in neighbour number, but the mean is still $6$. The Delaunay triangulation (DT) corresponding to a VT gives the appropriate graph connecting neighbouring cells. See Section~\ref{sec:VTmodel.model} and Figure~\ref{fig:VTandDT} for more detail on these terms. 

We measure the success of a cooperative mutant by comparing its fixation probability ($\rho_C$) to that of a neutral mutant ($\rho_0=1/N$). Thus if $\rho_C>1/N$ we say that cooperation is a beneficial mutation or that it is `favoured by selection'. The critical benefit-to-cost ratio, denoted $(b/c)^*$, is the point where the cooperator fixation probability is equal to the neutral fixation probability, i.e.\ $\rho_C=1/N$.

\begin{figure}[htb]
	\centering
	\includegraphics[width=0.6\textwidth]{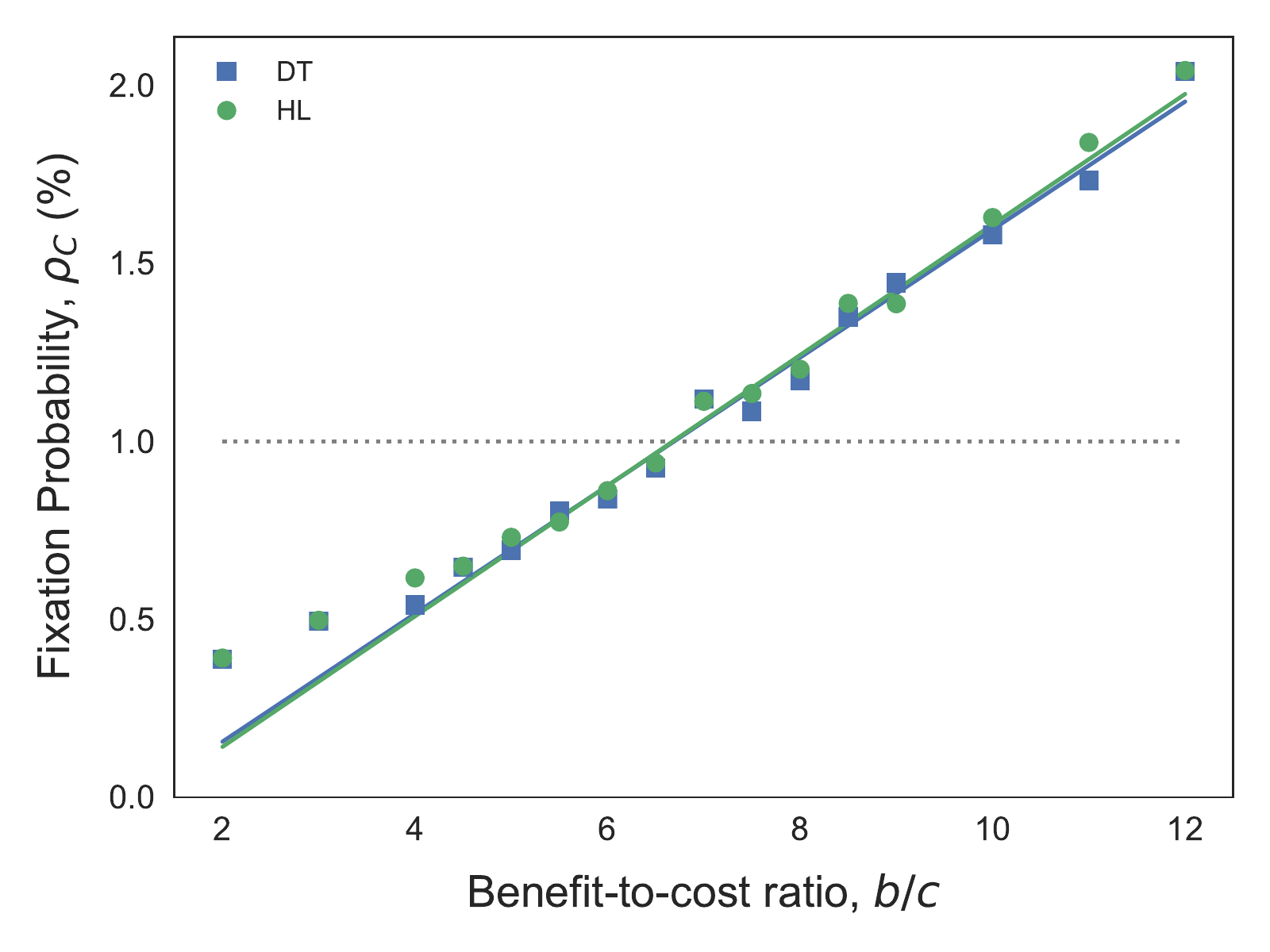}
	\caption{Fixation probabilities for a prisoner's dilemma game in the EGT model with $c=1$ and $\delta=0.025$. Solid lines plot theoretical fixation probabilities for a single cooperator on a hexagonal lattice (HL, green) and a Delaunay triangulation (DT, blue), obtained from Equation \eqref{eq:allen_fix}.  The critical benefit-to cost ratio, which occurs where fixation probability is equal to $\rho_0=1/N$ (grey), is $(b/c)^*\approx 6.7$ for the HL and DT. Simulation results are also shown for both cases and fit well with the theoretical fixation probabilites when $(b/c)>4$. However, as Equation~\eqref{eq:allen_fix} was derived in the weak selection limit we only expect it to be accurate near the critical ratio. }
	\label{fig:fixprob_allen}
\end{figure}

For a death-birth update rule we calculate the fixation probabilities against benefit-to-cost ratio $(b/c)$ for an HL and DT with a population size of $N=100$ and periodic boundary conditions. Results are plotted in Figure~\ref{fig:fixprob_allen} in which each data point is the result of \num{1e5} simulations. Analytical results are calculated using the theory developed in \cite{Allen2016}, where the authors derive an equation 
\begin{equation}
	\rho_C = \frac{1}{N}+\frac{\delta}{2N}\left(-ct_2+b(t_3-t_1)\right) +\mathcal{O}(\delta^2) \label{eq:allen_fix}
\end{equation} for the fixation probabilities on any graph. Here $t_n$ is the expected coalescence time from the two ends of an $n$-step random walk, where the initial vertex is chosen proportional to degree. Thus these quantities are purely properties of the graph and can be calculated computationally by solving a recurrence relation. We use a small selection strength, $\delta=0.025$, and there is a good fit between simulation and theory in the range shown for $b > 4$. Furthermore the heterogeneity in the DT seems to have a negligible effect on fixation probabilities compared to the dependence on benefit-to-cost ratio. These are calculated for both graphs from simulations and Equation~\eqref{eq:allen_fix} and summarised in Table~\ref{table:critical_ratios}.

\begin{table}[htb!]
	\centering
	\begin{tabular}{|l||c|c|}
		\hline
		\textbf{} & Theory & Simulation\\
		\hline \hline
		             EGT model with DT (death-birth) &              6.69 &              6.74\\ \hline
		             EGT model with HL (death-birth)&              6.68 &              6.67\\ \hline
		             VT model (decoupled update) &              2.78 &              2.83\\ \hline
					 VT model (death-birth) &              -   &           7.26\\ \hline
	\end{tabular}
		\caption{Summary of critical benefit-to-cost ratios, $(b/c)^*$, for the different models: a Moran process with death-birth update on a Voronoi network and a hexagonal lattice; a Voronoi tessellation model. Results are shown for both the theory and simulations.}
		\label{table:critical_ratios}
\end{table}

The results are very different for a birth-death update rule: cooperation is never favoured by selection under an additive prisoner's dilemma game and $\rho_C<1/N$ for all $b<c$, $c>0$ \cite{Ohtsuki2006,Taylor2007,Zukewich2013}. Thus within the EGT framework cooperation is only a successful evolutionary strategy on an epithelial structure with a death-birth update above a critical benefit-to-cost ratio of approximately $6.7$. 

The HL seems to be a reasonable approximation to the structure. Using the more realistic DT with neighbour number heterogeneity does not significantly alter fixation probabilities or the critical benefit-to-cost ratio, at least in the weak selection limit we are using. We note however that these results are for an average payoff and that an accumulative payoff (in which payoffs are simply summed over interactions) can amplify differences due to heterogeneity. We should also note that cooperation is possible in well-mixed populations or graph structured populations with birth-death update for games other than the prisoner's dilemma, such as the snowdrift or stag-hunt games, and it is possible to generalise \eqref{eq:allen_fix} to analyse these \cite{Allen2016}. 

Whether or not these results are illuminating in terms of a real epithelium is an important question however, and as we have noted previously there are some serious shortcomings to the model, first that population structure is static and secondly the troubling dependence on the update rule. Which update rule is closest to reality is unclear and while there likely is some coupling in birth and death processes in a real epithelium, there is certainly no absolute requirement for birth and death events to occur next to each other. In order to explore whether these factors are important to the dynamics we will move on to consider the VT model of an epithelium in which cells are able to move past each other and birth and death are spatially decoupled.


\section{Voronoi tessellation model of an epithelium}
\label{sec:VTmodel}
In order to analyse the dynamics of evolutionary games on a more realistic population structure we will use the VT model \cite{Meineke2001,VanLeeuwen2009} developed for the colonic crypt epithelium. In the following we will explain how the mechanical model works and generates a time-dependent graph structure on which to study evolutionary game dynamics. We will then derive an approximation for the fixation probability and use these results along with simulation to compare with the EGT model.

\subsection{The model} \label{sec:VTmodel.model}
The VT model represents a tissue as a set of points corresponding to the centres of individual cells. These points lie in a fixed domain with periodic boundary conditions. Cells move freely in space and exert spring-like forces on one another, such that 
\begin{equation}
	\mathbf{F}_{ij}(t) = \mu\hat{\mathbf{r}}_{ij}(t)(\vert \mathbf{r}_{ij}(t) \vert -s_{ij}(t)) 
\end{equation}
is the force exerted by cell $j$ on its neighbour $i$. Here $\mu$ is the spring constant and $\mathbf{r}_{ij}=\mathbf{r}_{i}-\mathbf{r}_ {j}$, where $\mathbf{r}_i$ is the position vector of cell $i$ and $\hat{\mathbf{r}}_{ij}$ is the corresponding unit vector. The natural seperation between cells $s_{ij}(t)=s$ is constant and the same for all neighbour pairs. The exception to this is for newborn sister cells for whom $s_{ij}$ grows linearly from $\epsilon$ to $s$ over the course of an hour. 

The total force acting on cell $i$ is then 
\begin{equation}
	\mathbf{F}_i(t)= \sum_{j \in \mathcal{N}_i(t)}\mathbf{F}_{ij} \; ,
\label{eq:force_i} \end{equation}
where $\mathcal{N}_i(t)$ is the set of cells neighbouring $i$. By assuming that motion is over-damped due to high levels of friction we obtain the equation of motion for each cell in the form of a first order differential equation 
\begin{equation}
	\eta \frac{d\mathbf{r}_i}{dt}= \mathbf{F}_i(t) \; ,
\end{equation}
where $\eta$ is the damping constant. This is solved numerically using 
\begin{equation}
	\mathbf{r}_i(t+\Delta t)= \mathbf{r}_i(t)+\frac{\Delta t}{\eta} \mathbf{F}_i \; ,
\end{equation}
where $\Delta t$ is a sufficiently small time step for numerical stability. For the parameter values used in our simulations see Table~\ref{tab:VTmodel_params}. 

\begin{table}[htb!]
	\centering
	\begin{tabular}{|c|l|l|}
		\hline
		\textbf{Parameter} & \textbf{Description} & \textbf{Value} \\
		\hline \hline
		             $\mu$ &  Spring constant                         &   50.0  \\
		               $s$ &  Natural seperation of mature cells      &   1.0   \\
			    $\epsilon$ &  Initial seperation of sister cells      &   0.05  \\
		            $\eta$ &  Drag coefficient                        &   1.0   \\
		        $\Delta t$ &  Time step (h)                           &   0.005 \\
				$\lambda$  &  Division and apoptosis rate  (h$^{-1}$) &   $12.0^{-1}$  \\
		\hline                                            
	\end{tabular}
		\caption{Table of parameters used in the Voronoi tessellation model \cite{Osborne2017}.}
			\label{tab:VTmodel_params}
\end{table}

\begin{figure}
	\centering
	\setlength{\fboxsep}{0pt}
	\setlength{\fboxrule}{0.7pt}
	\fbox{\includegraphics[width=0.45\linewidth]{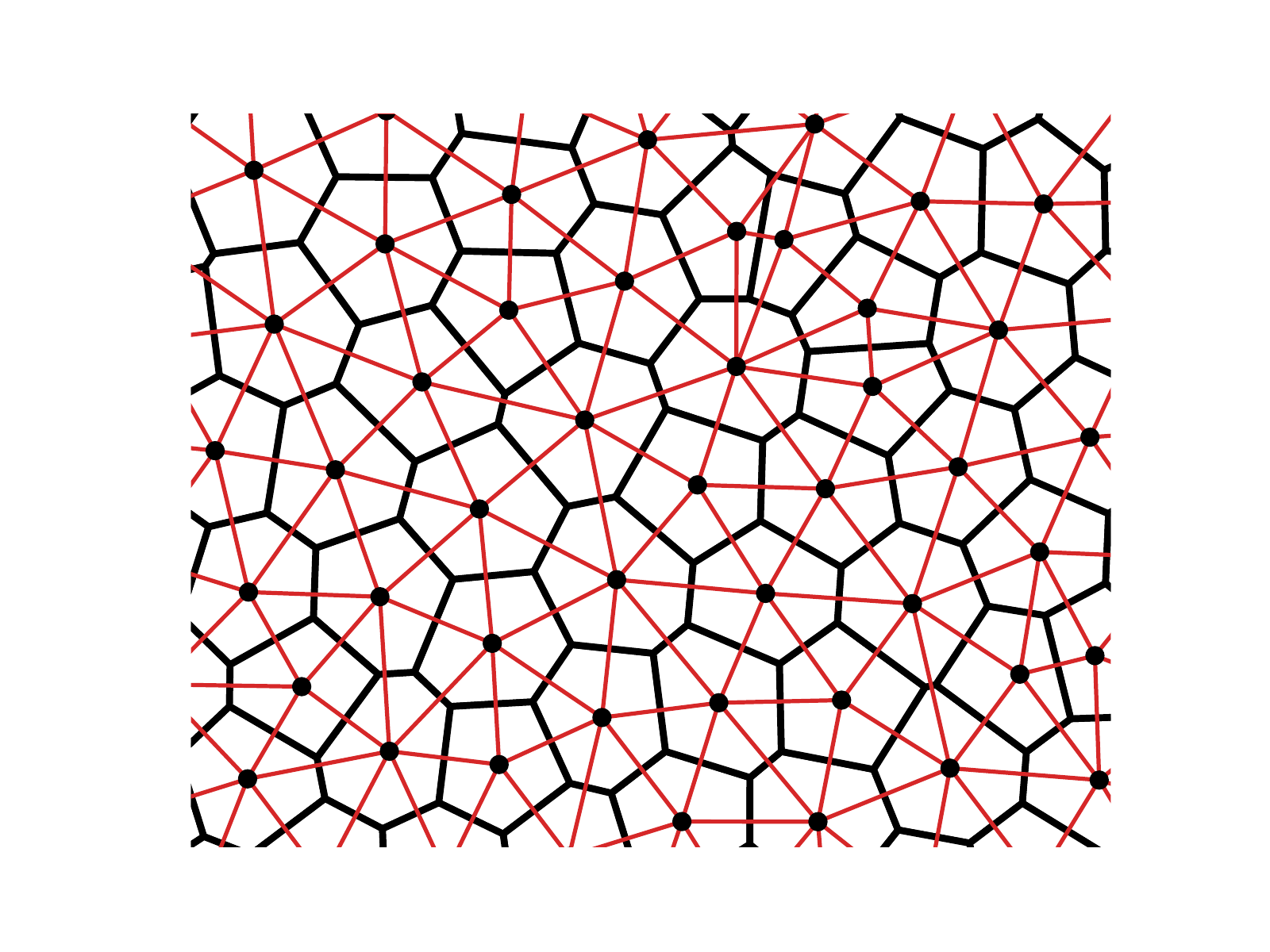}}
	\caption{Voronoi tessellation (VT, black) and Delaunay triangulation (DT, red) of a set of points representing cell-centres. The VT divides the plane into polygons such that every point in a polygon is closer to its corresponding cell-centre than any other. The DT partitions the plane into triangles and is the dual graph to the VT. Spring forces act along the lines of the DT }
	\label{fig:VTandDT}
	
\end{figure}

The neighbour connections between cells are determined by the VT of the set of cell-centres (see Figure~\ref{fig:VTandDT}). The VT divides the plane into polygons, where each polygon is defined as the the region of the plane closer to its generator (i.e.\ cell-centre) than any other. Each cell can therefore be represented as a distinct region with a well-defined area and neighbour set. The dual graph to the VT is the Delaunay triangulation (DT) in which the cell centres are the graph vertices and neighbours are connected by edges. The DT therefore gives the adjacency matrix $A_{ij}(t)$ from which we can calculate cell fitnesses. As it is defined by the cell-centre positions, the DT must be recalculated after every timestep during which cells may have moved, died or reproduced.

\begin{figure}
	\centering
	\includegraphics[width=0.7\linewidth]{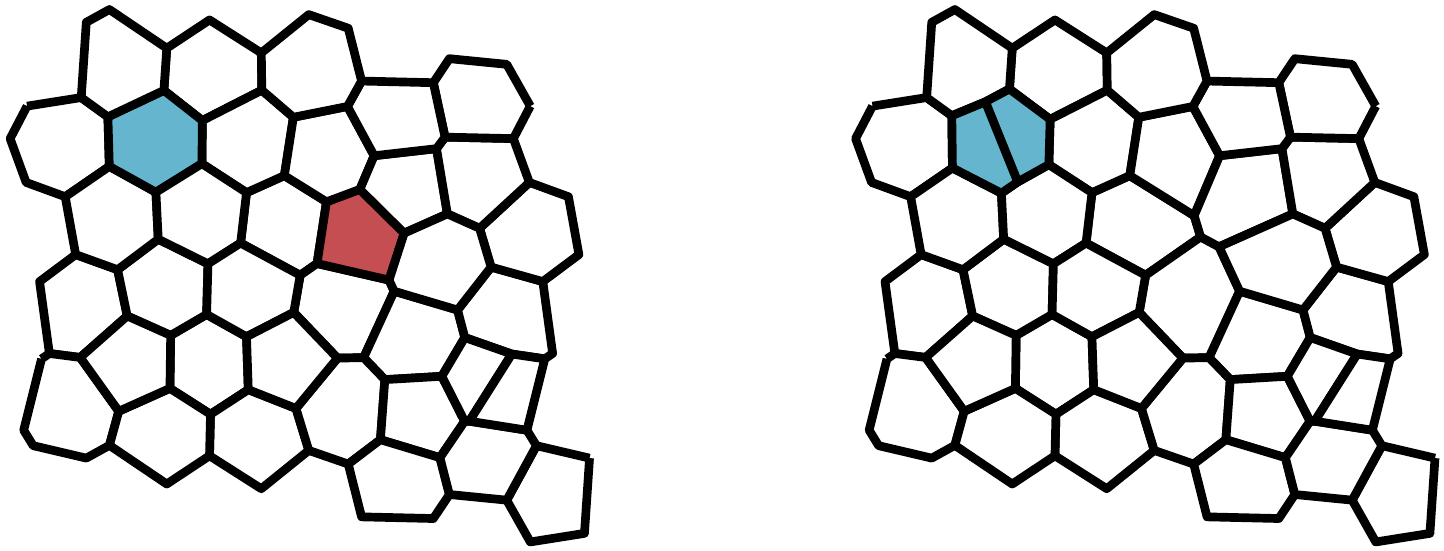}
	\caption{Spatially decoupled update rule in the Voronoi tessellation model. When an update event occurs a mother cell is chosen to reproduce with probability proportional to fitness (blue). A second cell is chosen to die uniformly at random (red). The mother cell divides and the dead cell is removed from the tissue.}
	\label{fig:vt_update}
	
\end{figure}

As in the previous model we allow the system to evolve by a Moran process whereby birth and death events occur simultaneously. The key difference is that we decouple the locations of these events. We also implement the process in continuous rather than discrete time, noting that a translation to continuous time in the previous model does not affect fixation probabilities \cite{Allen2016} and therefore the results are directly comparable. In the continuous time Moran process update events occur at exponentially distributed times with rate $\lambda$. When an update event occurs a mother cell is chosen at random from the population with probability proportional to fitness. This cell divides creating two daughter cells, which are exact clones of the mother. A cell is also chosen to die (i.e.\ to be extruded from the tissue) uniformly at random. This process is represented in Figure~\ref{fig:vt_update}.

To calculate fixation probabilities for a single mutant cooperator invading a defector population in the VT model we run simulations as follows. We begin with defector cells placed on a regular hexagonal lattice with periodic boundary conditions and the simulation algorithm proceeds until the system has relaxed into a dynamic equilibrium. We then choose a random cell to become a cooperator and continue the simulation until only cooperators or defectors remain. The simulation algorithm consists of the following steps: (1) DT is performed to determine cell neighbours; (2) forces are calculated and the cells moved accordingly; (3) an update event occurs with probability $N\lambda\Delta t$, in which case fitnesses are calculated according to the evolutionary game and the decoupled update rule is applied.

\subsection{Approximating the fixation probabilities} 
\label{sec:fix_prob_VT}
Due to the complexities of the VT model it is not possible to derive exact analytical solutions as was done for EGT \cite{Allen2016}. Instead we look for approximate solutions by considering the expected fitness for different cell types in populations with a given number of cooperators \cite{Pinheiro2012}. While the graph is dynamic and dependent on the spatial distribution of points, it is also planar and mechanically constrained. Furthermore if we begin with a single mutated cell, its progeny are likely to remain in a cluster as the clone grows. Thus we assume that variation in fitnesses for cells of each type will be small for a given number of cooperators in the population and that the average over a large number of states is a good approximation. Comparing our theoretical results to simulations we find that fixation probabilities calculated based on this assumption are good approximations.

\begin{figure}
	[htbp] \centering 
	\includegraphics[width=0.7
	\textwidth]{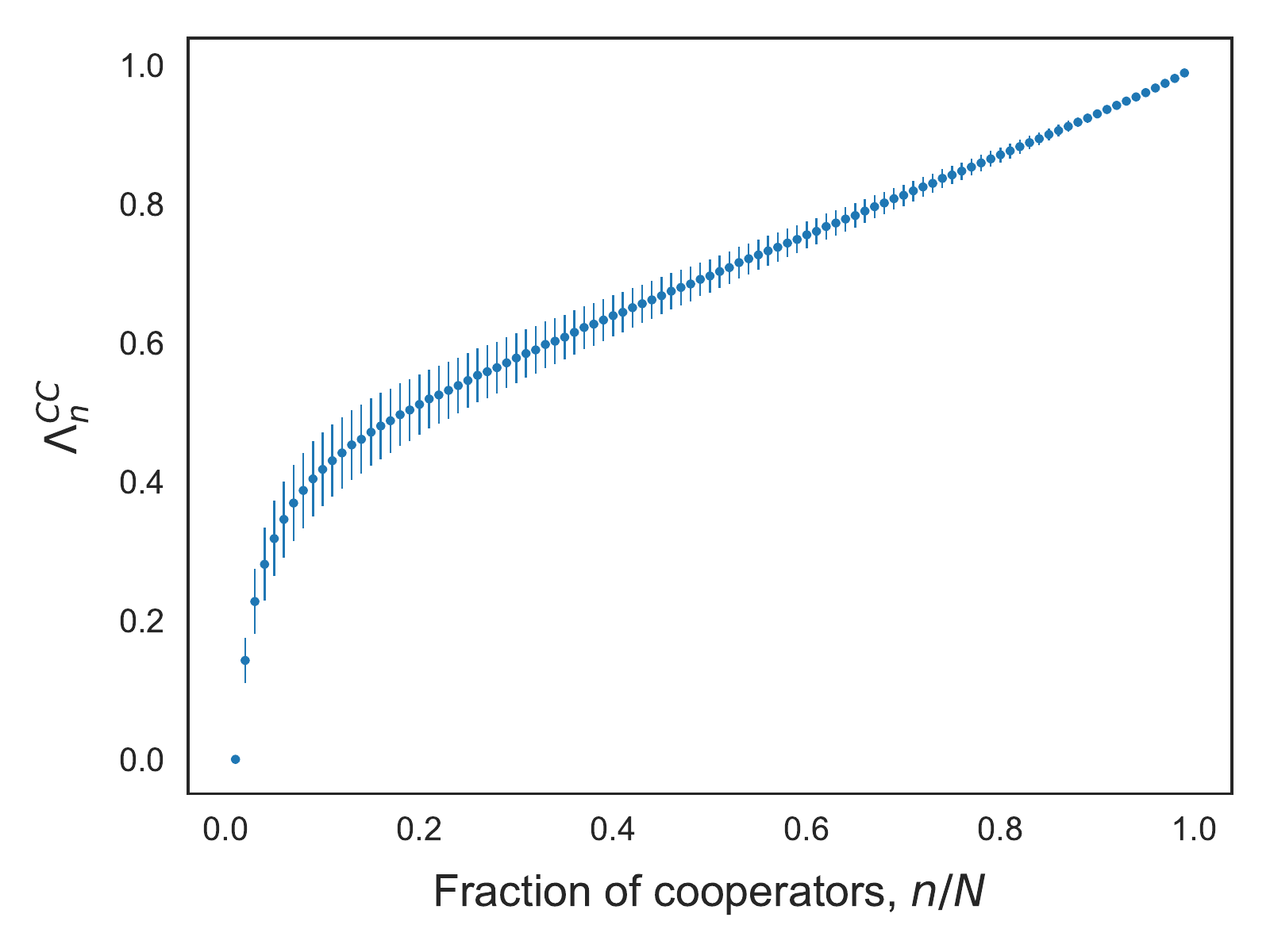}
	\caption{We calculate $\Lambda^{CC}_n$ for $N=100$ by running simulations of the VT model and tracking clones. In each simulation we look at snapshots in time give us a potential `state' from which to find the total number of degree-weighted cooperator-cooperator interactions for different clone sizes. This is then normalised and $\Lambda^{CC}_n$ is calculated by taking the mean over at least 5000 values. Error bars show standard deviation. } 
\label{fig:Lambda_CC} \end{figure}

Let us denote a state with $n$ cooperators $S_n=(\mathbf{s}_n,G)$, where $\mathbf{s}_n$ is the vector of cell types and $G$ is the graph. Then we define $T^{+/-}(S_n)$ to be the probability that when an event occurs the number of cooperators is increased/decreased by one, i.e\
\begin{align}
	T^+(S_n) &= \left(1-\frac{n}{N}\right)\frac{\sum_{i\in G}s_iF_i}{\sum_{i\in G}F_i} \label{eq:T+_state_full}\\
	T^-(S_n) &= \frac{n}{N}\left(1-\frac{\sum_{i\in G}s_iF_i}{\sum_{i\in G}F_i}\right) \, .
\label{eq:T-_state_full} \end{align}
We can then define the average transition probabilities for a state with $n$ cooperators to be $T^\pm_n=\langle T^\pm(S_n)\rangle$ where the average is taken over a large ensemble of possible states. Substituting in for the fitnesses \eqref{eq:fitness} and taking the weak selection limit $\delta\ll 1$ we obtain
\begin{align}
	T^+_n&= \frac{n}{N}\frac{N-n}{N}\left(1+\delta\langle f_C -f \rangle_0\right) + \mathcal{O}(\delta^2) \label{eq:T+}\\
	T^-_n&=\frac{n}{N}\frac{N-n}{N}\left(1-\frac{n}{n-N}\delta\langle f_C-f\rangle_0\right)  + \mathcal{O}(\delta^2) \label{eq:T-} \; ,
\end{align}
where $\langle .\rangle_0$ denotes an average over a large ensemble of possible states for the neutral process $\delta=0$ and 
\begin{align}
	f_C = \frac{1}{n}\sum_{i\in G}s_if_i && f = \frac{1}{N}\sum_{i\in G}f_i 
	\label{eq:avpay}
\end{align}
are the average cooperator fitness and average fitness respectively. From \eqref{eq:payoffs} and \eqref{eq:avpay} we obtain
\begin{equation}
	\langle f_C-f\rangle_0 = -c\left(1-\frac{n}{N}\right) +b\left(\Lambda^{CC}_n -\frac{n}{N}\right) \; ,
\end{equation} 
where 
\begin{equation}
	\Lambda^{CC}_n=\frac{1}{n}\left\langle\sum_{i,j \in G}\frac{s_is_jA_{ij}}{k_i} \right \rangle_0 
\end{equation}
is the normalised average number of degree-weighted cooperator-cooperator interactions in a system with $n$ cooperators. This can be calculated computationally by running simulations for a neutral process and tracking clones (groups of cells with common ancestry). At each time interval we calculate the contribution to $\Lambda^{CC}_n$ for all clones in the system, treating each lineage as a group of $n$ cooperators in a population of defectors. See Figure~\ref{fig:Lambda_CC} for a plot of $\Lambda^{CC}_n$ with $N=100$.

We use the equation for cooperator fixation probability derived in \cite{Taylor2004} for a well-mixed population
\begin{equation}
	\rho_C = \left[1+\sum_{m=1}^{N-1}\prod_{n=1}^{m}\gamma_n\right]
\end{equation}
with $\gamma_n = T^-_n/T^+_n$. In that case the transition probabilities and thus $\gamma_n$ are defined exactly for each value of $n$. For the VT model we substitute in the mean transition probabilities given by Equations \eqref{eq:T+},~\eqref{eq:T-} and \eqref{eq:avpay}, to obtain
\begin{equation}
	\rho_C \approx \frac{1}{N}+\frac{\delta}{N}\left\{\frac{-c(N-1)}{2}+b\sum_{m=1}^{N-1}\sum_{n=1}^{m}\left(\frac{\Lambda^{CC}_n-n/N}{N-n}\right)\right\} +\mathcal{O}(\delta^2)\label{eq:vtmodel_fp}
\end{equation}
for the fixation probability in the weak selection limit.
 The critical benefit-to-cost ratio is then obtained by setting $\rho_C=1/N$ giving
\begin{equation}
	\left(\frac{b}{c}\right)^*\approx\frac{N-1}{2}\left[\sum_{m=1}^{N-1}\sum_{n=1}^{m}\left(\frac{\Lambda^{CC}_n-n/N}{N-n}\right)\right]^{-1} \, . \label{eq:ratio_vt}
\end{equation}

\begin{figure}
	[htbp] \centering 
	\includegraphics[width=0.7
	\textwidth]{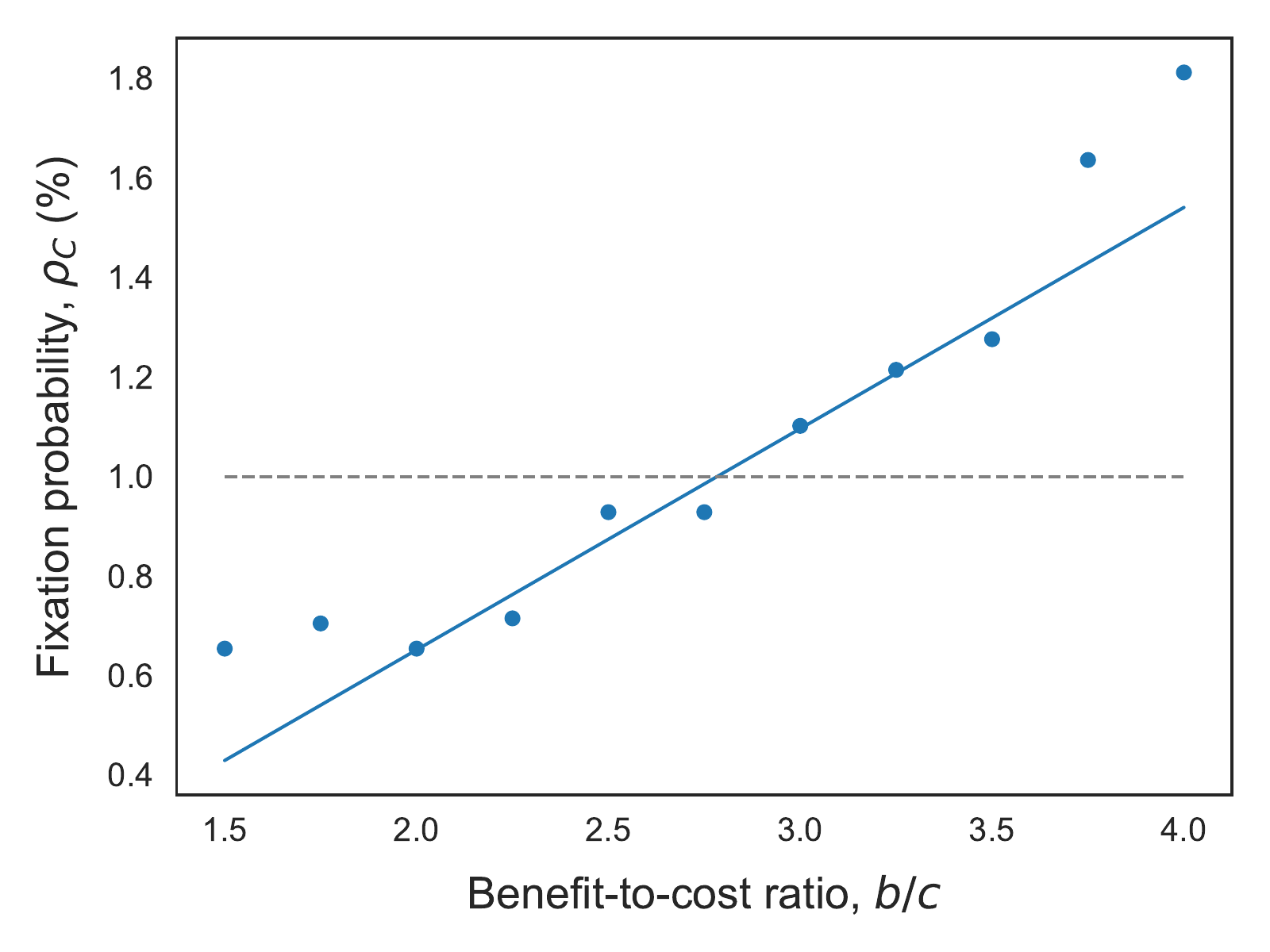} 
	\caption{An approximation for fixation probabilities in the VT model is given by Equation~\eqref{eq:vtmodel_fp} and plotted here (solid line) for $\delta=0.025$, $c=1$ and $N=100$. Comparison with simulation results (points) shows that the approximation is good near the critical benefit-to-cost ratio (i.e. where $\rho_C=\rho_0=1/N$), but breaks down outside the region $2<b<3.5$. This is consistent with the fact that the equation is derived in the weak selection limit, and suggests that it can be used to calculate the critical ratio.} 
\label{fig:fixprobs1} \end{figure}

Figure~\ref{fig:fixprobs1} compares Equation~\eqref{eq:vtmodel_fp} with simulation results for the VT model. It shows there is a reasonable fit between our approximation of fixation probabilities with the simulation data in the region $2.0<b<3.5$, where we have once again set $c=1$. These values are close to the critical benefit-to-cost ratio and therefore represent the region in which we would expect the weak selection limit to hold, thus this equation for fixation probabilities is a reasonable approximation. The critical benefit-to-cost ratios calculated from simulation and Equation~\eqref{eq:ratio_vt} are given in Table~\ref{table:critical_ratios}. For both we get a value of $b/c=2.8$ correct to one decimal place. This is significantly less than the critical benefit-to-cost ratios calculated for the EGT model with death-birth update. In the next section we will look further at comparing these models and attempt to identify the mechanism by which cooperation is promoted in the VT model.

\section{Comparing the models}
\label{sec:comparing_the_models}

\begin{figure}[htb]
	\centering
	\includegraphics[width=0.7\textwidth]{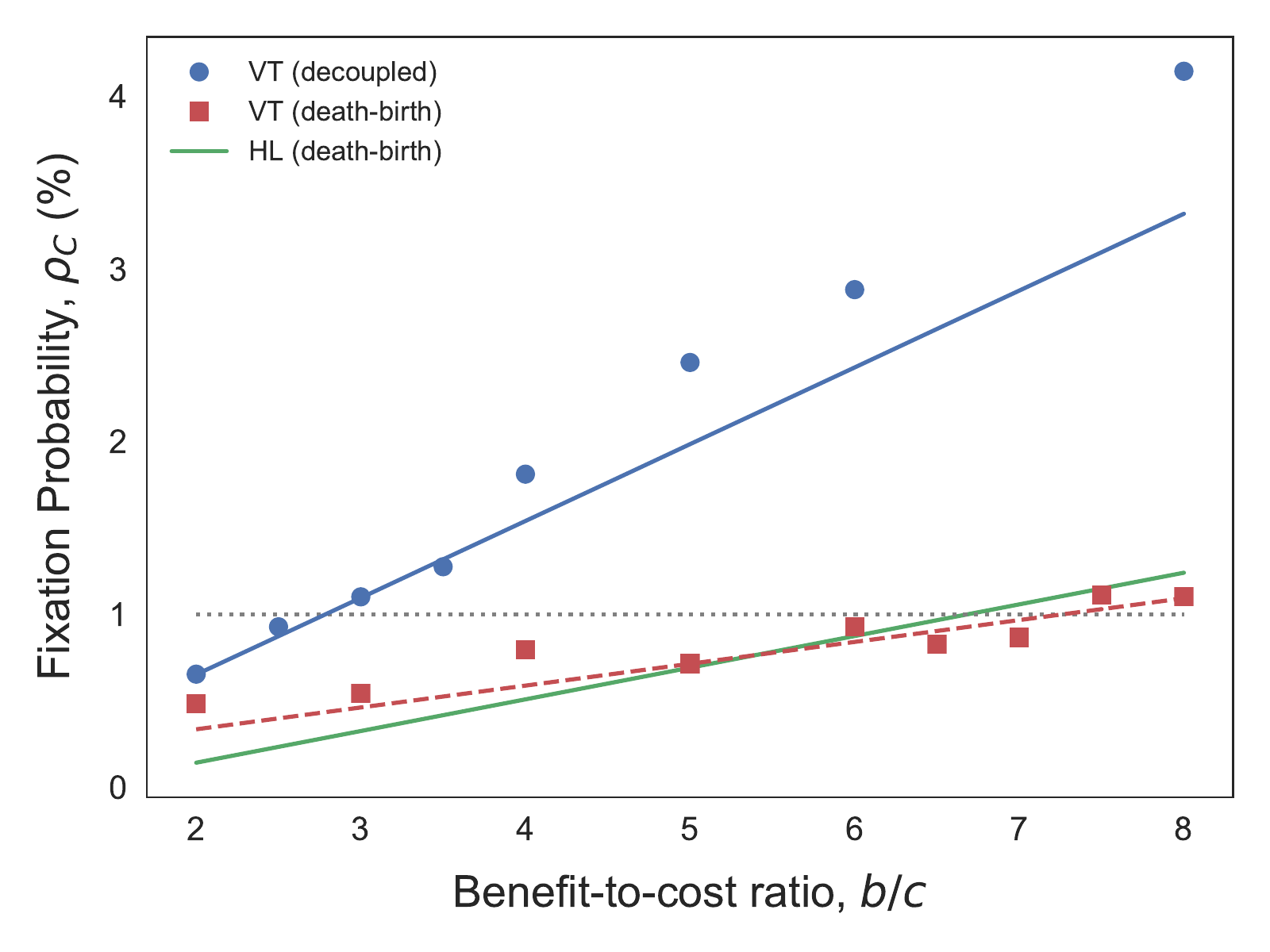}
	\caption{Fixation probabilities for the prisoner's dilemma game in the VT model with $c=1$ and $\delta=0.025$. Points show simulation results for a decoupled update rule (blue, circles) and a death-birth update rule (red, diamonds).  For the decoupled update rule the approximate fixation probabilities given by Equation~\eqref{eq:vtmodel_fp} are plotted (blue, solid line) and for the death-birth update we plot a best fit line (red, dashed line). Fixation probabilities, given by Equation~\eqref{eq:allen_fix}, for an HL with death-birth update in the EGT model (green, solid line) are also shown for comparison. 
	The grey dotted line shows the fixation probability for a neutral mutant. It is clear that cooperation is significantly favoured in the VT model with decoupled update rule when compared with the EGT results, in particular the critical benefit-to-cost ratio is more than halved. However when a death-birth update is introduced on the VT model this effect is suppressed and the critical benefit-to-cost ratio is very close to the EGT case.}
	\label{fig:vtcompare}
\end{figure}

\begin{figure}[htb]
	\centering
	\includegraphics[width=0.7\textwidth]{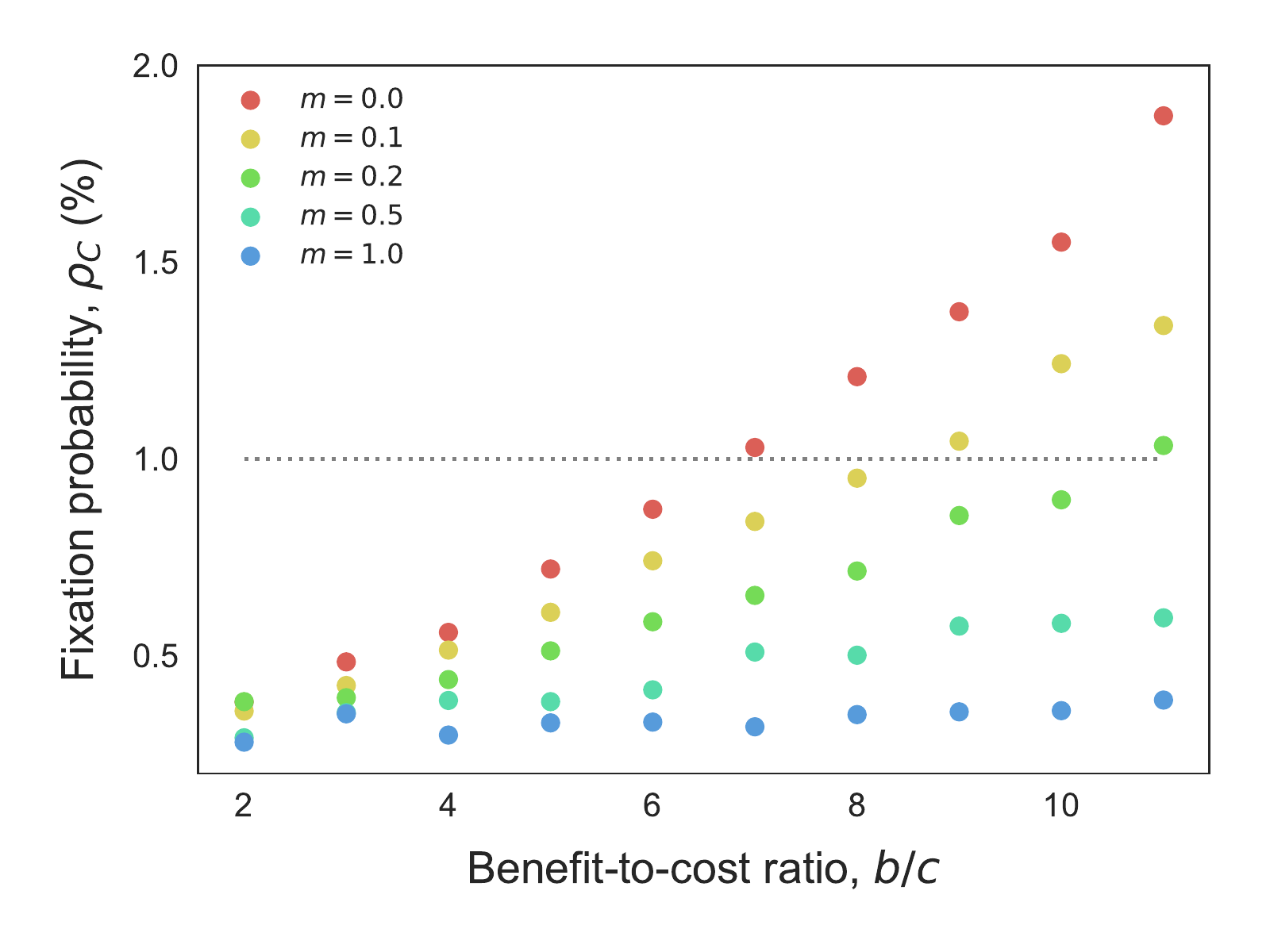}
	\caption{Fixation probabilities for an EGT model with migration on an HL, with $c=1$ and $\delta=0.025$, are obtained through simulation. The parameter $m$ is the probability that a migration event will occur in each timestep. If such an event occurs two neighbouring cells are randomly selected to swap vertices. Increasing $m$ leads to decreased cooperative success. The case $m=0$ corresponds to our original EGT model with no migration. }
	\label{fig:hex_migration}
\end{figure}

Figure~\ref{fig:vtcompare} shows the results of these simulations along with the theoretical EGT results for the HL graph with death-birth update and the critical benefit-to-cost ratios are summarised in Table~\ref{table:critical_ratios}. It is clear that cooperators are much more successful in the VT model, in particular the critical benefit-to-cost ratio for the VT model is less than half that for EGT with death-birth update. The question then arises as to what mechanism is causing this amplifying effect in the VT model, the two obvious candidates being the effect of cell motility and the decoupling of birth and death. One way to test whether cell motility is enhancing the evolutionary success of cooperation is to introduce an analogue into the EGT model whereby we allow cells to swap sites with their neighbours. At each timestep a swap occurs with probability $m$. When this happens a cell is chosen uniformly at random to switch places with one of its neighbours. Note that this process is independent of cell fitness. The parameter $m$ is therefore a measure of the strength of migration and by setting $m=0$ we regain the original EGT model. Figure~\ref{fig:hex_migration} plots fixation probability against benefit-to-cost ratio for a range of $m$ values and demonstrates that increasing migration within this framework actually decreases the evolutionary success of cooperation. It therefore seems unlikely that the ability of cells to move past each other in the VT model is the reason for enhanced cooperative success. 

\begin{figure}
	\centering
	\includegraphics[width=0.7\textwidth]{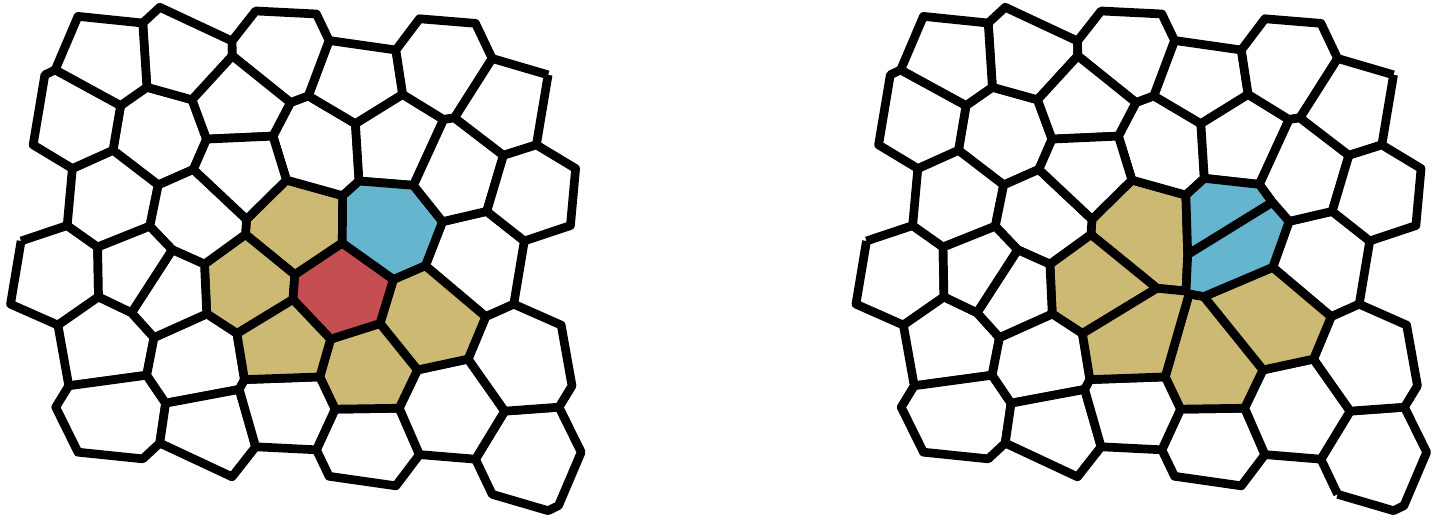}
	\caption{Death-birth update rule in the Voronoi tessellation model. When an update event occurs a cell is chosen to die uniformly at random from the population (red). From the neighbourhood of the dead cell (yellow) a mother cell (blue) is then chosen with probability proportional to fitness. The mother cell divides and the dead cell is removed from the tissue.}
	\label{fig:db_on_VT}
\end{figure}

In order to determine whether the spatial decoupling of birth and death promotes cooperation we consider the VT model with a death-birth update rule. To implement this we follow the simulation algorithm as defined in Section~\ref{sec:VTmodel}, the only change being in choosing which cells reproduce and die when an update event occurs. First a cell is chosen for extrusion uniformly at random. Fitnesses are then calculated for the neighbouring cells and one of these is chosen to divide with probability proportional to fitness. This process is shown schematically in Figure~\ref{fig:db_on_VT}. It can be seen clearly in Figure~\ref{fig:vtcompare} that changing the update rule in this way suppresses the evolutionary success of cooperation in comparison to the decoupled update rule. Indeed in this case we obtain $b/c=7.3$ which is greater than for the EGT model with death-birth update. 

Combining these two results we conclude that it is the spatial decoupling of birth and death which leads to the amplification of cooperative success in the VT model. Indeed this is an intuitive result and is consistent with results from the shift dynamics models \cite{Allen2012,Pavlogiannis2015}. A cooperative strategy is only beneficial if cells are able to form a cluster of cooperators. If birth and death are constrained to occur next to each other, as is the case for death-birth and birth-death update rules, then the cluster can only grow at the boundary. If a cell were to reproduce inside a cooperative cluster it would result in the death of a neighbouring cooperator, leaving the size of the cooperator population unchanged. For the decoupled birth and death update in the VT model this is not the case. If a cooperator inside the cluster reproduces it will lead to an increase in the size of the cooperator population with probability $1-n/N$, where $n$ is the number of cooperators and $N$ the total number of cells. The fact that migration appears to suppress the success of cooperation could also provide an explanation as to why, if a death-birth update is enforced in both cases, cooperators fare better in the EGT model than in the VT model.

\section{Conclusions} 
\label{sec:conclusions}

Evolutionary graph theory has become the accepted framework for modelling the evolution of cooperation on structured populations, ranging from complex social networks to collective cellular behaviour organised in tissues. While it may be an appropriate tool for the former, we have demonstrated that a static graph model is not sufficient to capture the dynamic behaviour of an epithelium. 

We have shown using the theory developed by Allen et al. \cite{Allen2016} and simulations that for a prisoner's dilemma on an epithelium-like structure in EGT, cooperation is successful if $b/c > 6.7$ for a death-birth update, where we have used an averaged payoff. This inequality holds when we model the epithelium as an HL as well as a DT, suggesting that there is a marginal effect on fixation probabilities due to heterogeneity of neighbour number. However, the choice of an averaged payoff could be suppressing the effect of heterogeneity compared to an accumulated payoff, as it does for scale-free networks \cite{Szolnoki2008,Maciejewski2014a}. It would be advisable therefore to compare fixation probabilities on the two structures for an accumulated payoff, although we do not expect a substantial difference. Vertex degree in scale-free networks follow a polynomial distribution and therefore exhibit large variance, whereas degree variance in DTs is comparatively small.

For a birth-death update on the other hand, cooperation is not successful for any benefit-to-cost ratio under a prisoner's dilemma game. The fact that the dynamics is so sensitive to the choice of update rule is troubling and neither update rule is a realistic representation of birth and death on an epithelium. For the VT model we are able to spatially decouple birth and death. We showed, using simulation and approximate theoretical results, that using a decoupled update rule in the VT model promotes cooperation compared to the EGT examples. Furthermore when the VT model was run with a death-birth update this effect was suppressed and cooperation actually fared worse than in the EGT model, leading us to conclude that the decoupling of birth and death is the main mechanism for increased success of cooperation in the VT model. This is consistent with previous work looking at shift dynamics on a static graph which found that decoupling birth and death led to increased cooperative success in one-dimension \cite{Allen2012}, and in two-dimensions if a repulsive force was introduced between cells of different types \cite{Pavlogiannis2015}. The fact that cells can move and change neighbours in the VT model however, does not appear to increase the likelihood of cooperation fixating. Indeed we found that introducing migration into an EGT model actually suppressed cooporation, and it is therefore possible that cell motility is acting to reduce cooperative success in the VT model.

As it is the update rule which seems to influence the evolutionary success of cooperation most substantively, the question arises as to which, if any, reflects the behaviour of a real epithelium. Clearly it is unrealistic that when a death occurs it is immediately followed by a neighbour undergoing division, or vice versa, as for the death-birth and birth-death update rules respectively. However it is also not the case that birth and death events are completely spatially independent. Cell extrusion can be induced in areas of overcrowding within a tissue, which could be caused by high levels of proliferation. Similarly if local density is low, e.g.\ due to a high instance of cell death, cells can be induced to reproduce \cite{Eisenhoffer2012,Bove2017}. It is difficult to see how this more subtle link between birth and death could be implemented in an EGT model, however the VT model could be extended to include density-dependence for division and/or extrusion. Furthermore a density-dependent model would allow us to maintain an (almost) constant population size without enforcing that birth and death occur simultaneously, another unrealistic assumption.

In our discussion of whether cooperation is succesful on an epithelium we have limited ourselves to the additive prisoner's dilemma game, whereas evolutionary game theory models of cancer have used a variety of social dilemma games. Extending our analysis to a general two-strategy game should be relatively straightforward, indeed we can use the critical benefit-to-cost ratio to calculate the structure coefficient and derive a general condition for evolutionary success for a two-player, two-strategy game \cite{Tarnita2009}. However it has been argued that multiplayer public goods games are more realistic for cancer modelling, and can lead to very different results. Recent work has considered the dynamics of these types of games on lattices \cite{Archetti2013a} and DT graphs \cite{Archetti2016} in an EGT framework, it would therefore be an interesting comparison, but non-trivial extension, to consider them on the VT model.

\vspace{10pt}
\begin{small}
\noindent
\textbf{Data accessibility.} The code and data can be accessed at https://github.com/jessiesrr/evo-epithelium.

\noindent 
\textbf{Authors' contributions.} JR and KMP designed the research. JR carried out the research and wrote the paper. KMP edited the paper.

\noindent 
\textbf{Acknowledgements.} We would like to thank John Talbot for helpful discussions and Pilar Guerrero and David Page for advice on the code. 

\noindent 
\textbf{Funding.} This research was funded by an EPSRC studentship held by JR.

\noindent 
\textbf{Competing interests.} We declare we have no competing interests.

\end{small}
\bibliographystyle{unsrtnat_edit} 
\bibliography{bib}

\end{document}